\documentclass[12pt,aps,preprint,nofootinbib,superscriptaddress,nobalancelastpage]{revtex4}

\usepackage{hyperref}
\usepackage{epsfig}
\usepackage{amssymb}
\usepackage{amsmath}
\usepackage{amsfonts}
\usepackage{graphics}
\usepackage{cancel}
\usepackage{bbm}
\usepackage{mathrsfs}
\usepackage{slashed}
\usepackage{color}
\usepackage{subfigure} 
\usepackage{multirow}

\newcommand{\bea}{\begin{eqnarray}}
\newcommand{\eea}{\end{eqnarray}}

\newcommand{\be}{\begin{equation}}
\newcommand{\ee}{\end{equation}}

\newcommand{\ba}{\begin{array}}
\newcommand{\ea}{\end{array}}

\allowdisplaybreaks

\begin{document}
\title{Physics reach of CERN-based SuperBeam neutrino oscillation experiments}

\author{Pilar Coloma}
\email[]{pcoloma@vt.edu}
\affiliation{Center for Neutrino Physics, Department of Physics, Virginia Tech, Blacksburg, VA 24061, USA}

\author{\mbox{Enrique~Fern\'andez-Mart\'inez}}
\email[]{enfmarti@cern.ch}
\affiliation{CERN Physics Department, Theory Division, CH-1211 Geneva 23, Switzerland}

\author{Luis Labarga}
\email[]{luis.labarga@uam.es}
\affiliation{Departamento de F\'isica Te\'orica, Universidad Aut\'onoma de Madrid, Cantoblanco
28049 Madrid, Spain}

\begin{abstract}
We compare the physics potential of two representative options for a SuperBeam in Europe, studying the achievable precision at $1 \sigma$ with which the CP violation phase ($\delta$) could be measured, as well as the mass hierarchy and CP violation discovery potentials. The first setup corresponds to a high energy beam aiming from CERN to a 100 kt liquid argon detector placed at the Pyh\"asalmi mine (2300 km), one of the LAGUNA candidate sites. The second setup corresponds to a much lower energy beam, aiming from CERN to a 500 kt water \v{C}erenkov detector placed at the Gran Sasso underground laboratory (730 km). This second option is also studied for a baseline of 650 km, corresponding to the LAGUNA candidate sites of Umbria and the Canfranc underground laboratory. All results are presented also for scenarios with statistics lowered by factors of 2, 4, 8 and 16 to study the possible reductions of flux, detector mass or running time allowed by the large value of $\theta_{13}$ recently measured.
\end{abstract}


\preprint{EURONU-WP6-12-51}

\maketitle

\section{Introduction}

Daya Bay~\cite{An:2012eh} and RENO~\cite{Ahn:2012nd} have recently confirmed the previous hints from T2K~\cite{Abe:2011sj}, MINOS~\cite{Adamson:2011qu}, Double-CHOOZ~\cite{Kuze:2011ic} and the interplay between solar and KamLAND data~\cite{GonzalezGarcia:2010er,Fogli:2011qn} with the discovery of a large value of $\theta_{13}$ which saturates previous upper bounds~\cite{Schwetz:2011zk}. Recent global fits~\cite{Tortola:2012te,Fogli:2012ua} give a best fit for $\theta_{13}$ between $\sin^2\theta_{13}=0.024$ and $0.027$ (with the larger values for an inverted hierarchy) and a $1 \sigma$ error close to a $10 \%$. Such a large value opens the window to fundamental measurements such as the existence of leptonic CP violation and the neutrino mass hierarchy, critical for a comparison with double neutrinoless beta decay searches probing the Majorana nature of the neutrino fields. The value of $\theta_{13}$ currently favoured would allow these searches to be performed at relatively modest upgrades of conventional neutrino beams to SuperBeam setups, characterized with a beam power close to (or above) 1~MW. In this work we will explore and compare the physics potential and performance of two representative setups for a European SuperBeam experiment with a neutrino flux produced at the CERN accelerator complex. 

Seven possible detector sites have been studied within the LAGUNA~\cite{laguna} project: Fr\'ejus (France), Canfranc (Spain), Umbria (Italy), Sierozsowice (Poland), Boulby (UK), Slanic (Romania) and Pyh\"asalmi (Finland). In addition there is the Gran Sasso (Italy) underground laboratory, which presently hosts the CNGS~\cite{cngs} physics program and is studying the only existing neutrino beam in Europe. Here we will concentrate in two extreme setups: the longest possible baseline of 2300~km corresponding to the distance from CERN to Pyh\"asalmi, and a shorter baseline of 730~km which corresponds to the present beamline between CERN to Gran Sasso. We will also discuss the physics performance of alternative LAGUNA sites with similar baselines to Gran Sasso such as Canfranc (650~km) or Umbria (665~km). 

An even shorter baseline of 130~km matching the CERN to Fr\'ejus distance has also been extensively studied~\cite{GomezCadenas:2001eu,Donini:2004hu,Campagne:2004wt,Donini:2004iv,Donini:2005db,Campagne:2006yx,Longhin:2011hn}. The low energies needed to match this short baseline imply correspondingly low cross sections and, typically, less statistics than other setups. If a high beam power around 4~MW is achievable in order to compensate the reduced cross section at these energies, this setup would provide an excellent sensitivity to leptonic CP violation, given the negligible matter effects that could mimic its presence. However, the small matter effects also imply no sensitivity to the mass hierarchy from the study of the oscillations of the neutrino beam alone, although some sensitivity can be gained in combination with atmospheric neutrino oscillations at the same detector~\cite{Huber:2005ep,Campagne:2006yx}. For the large values of $\theta_{13}$ currently favoured, an even more attractive option implies the observation of this low energy beam at its second oscillation peak, which would increase the CP violation discovery potential as well as the determination of the mass hierarchy, at a $\sim 650$~km baseline~\cite{Coloma:2011pg}. However, as these high beam powers are not expected to be achieved in the near future, in this work we will instead assume a more modest flux of $\sim 0.8$ MW, similar to what is being considered for LAGUNA-LBNO~\cite{andre}. For the high energy and long baseline option of 2300~km we will consider a 100~kt liquid argon (LAr) detector, while the lower energies required for the oscillation at 730~km match better the water \v{C}erenkov (WC) technology, for which we consider a 500~kt fiducial volume. In order to explore if the large value of $\theta_{13}$ allows for more conservative setups with reduced power, detector mass or running time, we will present all our results with reductions in the statistics by factors of 2, 4, 8 and 16.    

The paper is organized as follows. In Section~\ref{sec:setups} we introduce the experimental setups under study and the assumptions adopted to simulate their performance. In Section~\ref{sec:results} we show our comparison of the physics performance of the two setups for the precision on their measurement of the CP violating phase $\delta$, their CP violation discovery potential and their sensitivity to the mass hierarchy. Finally, in Section~\ref{sec:summary} we summarize and discuss the results and draw our conclusions in Section~\ref{sec:concl}.

\section{Setups}
\label{sec:setups}

\begin{figure}
\begin{center}
\includegraphics[width=1.\textwidth]{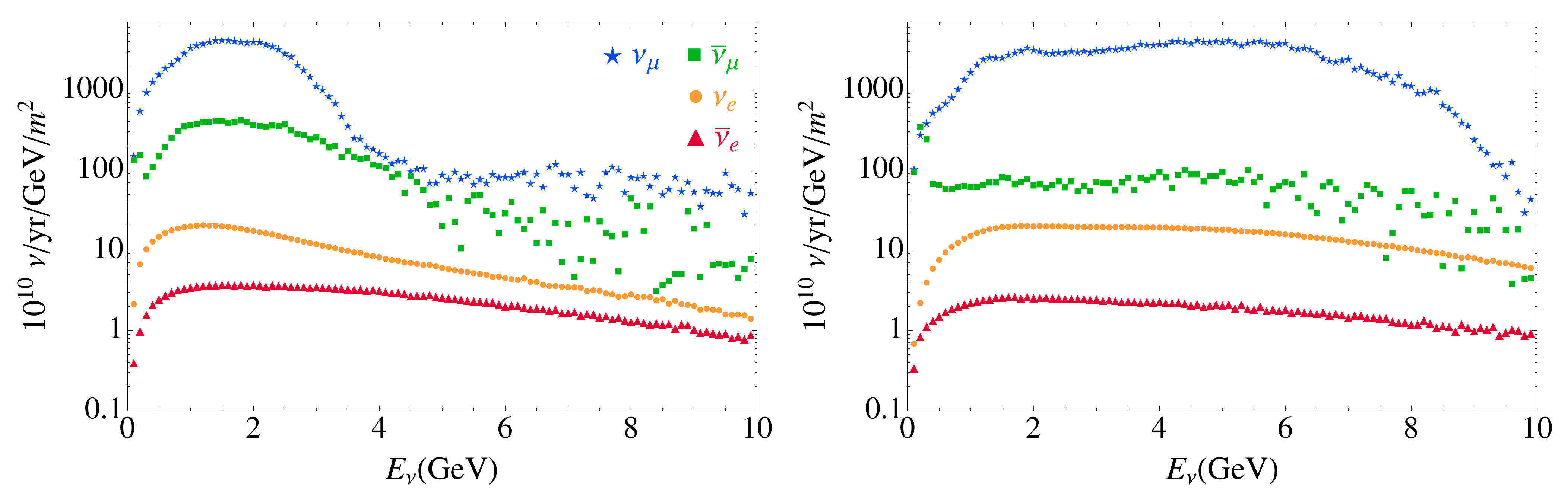} \\
\caption{ Neutrino fluxes for the two beam configurations that have been used in this paper. Left and right panels show the fluxes that have been used to simulate the results for the 730 and 2300 km baselines, respectively. The different symbols correspond to the main component of the beam and its intrinsic contamination in $\nu$ mode, as indicated in the legend. The composition of the beam in $\bar\nu$ mode is very similar. Fluxes have been taken from Ref.~\cite{longhin}, where $3\times 10^{21}$ PoT per year and 50 GeV protons were assumed. However, for the simulations presented in this paper we have reduced the number of PoT per year by a factor of three, corresponding to an integrated luminosity of 0.8 MW per year (assuming $10^7$ useful seconds).  }
\label{fig:fluxes}
\end{center}
\end{figure}

We will compare the physics performance of two CERN-based SuperBeam setups in combination with either a 100~kt LAr detector at 2300~km or a 500~kt WC detector at 730~km. To match these two baselines so as to have the oscillation probability roughly at the first oscillation peak we consider two different possible fluxes. A more energetic one with a mean neutrino energy around $\sim5$~GeV will be considered for the CERN-Pyh\"asalmi baseline, while a lower energy flux peaking around $\sim1.5$~GeV is better suited for the shorter 730~km baseline (see Fig.~\ref{fig:fluxes}). These fluxes were kindly provided by A.~Longhin and were computed for 50~GeV protons and $3 \cdot 10^{21}$ protons on target per year~\cite{longhin}, corresponding to the capabilities of an upgraded accelerator complex. For the analysis presented in this work we have decreased the number of protons on target by factor of 3, corresponding to a beam power of 0.8 MW per year (assuming $10^7$ useful seconds).

In order to simulate the WC detector response, we have followed the T2HK letter of intent~\cite{Abe:2011ts}. In particular we take the signal and background efficiencies from Tables VIII and IX in Ref.~\citep{Abe:2011ts}, for neutrino and antineutrino running modes respectively. Notice that, since the energy range for the considered flux is about a factor two higher than for the T2HK beam, a lower quasi-elastic (QE) event rate is expected, with a consequent reduced efficiency when the 1-ring cut is imposed compared to Tables VIII and IX of Ref.~\cite{Abe:2011ts}. In order to take this into account, we rescale all charged current efficiencies in those tables by removing the 1-ring cut but we only consider QE events for the charged current processes, which should constitute the dominant component that passes the 1-ring cut. This entails a 77\% (82\%) efficiency for the $\nu_e$ ($\bar{\nu}_e$) apperance channels. The background for the $\nu_e$ ($\bar\nu_e$) appearance channel is given by the full $\nu_e$ and $\bar\nu_e$ intrinsic contamination of the beam, plus a $0.06\%$ ($0.03\%$) of the $\nu_\mu$ events (which are misidentified as $\nu_e$) and a $1.0 \%$ ($1.3 \%$) of the neutral current events. Finally, efficiencies of $75.4\%$ and $68.1\%$ have also been considered for the oscillated background arising from the opposite polarity component of the beam ($\bar\nu_\mu \rightarrow \bar\nu_e$ and $\nu_\mu \rightarrow \nu_e$ for neutrino and antineutrino modes, respectively). We have assumed a 90\% efficiency for the $\nu_\mu$ and $\bar\nu_\mu$ disapperance channels with the same neutral current background contamination as for apperance. We have assumed these values to be constant in the neutrino energy range considered, between 0.4 and 4.4~GeV. A Gaussian energy resolution of 85~MeV was also considered as suggested by Fig.~2 of Ref.~\cite{Itow:2001ee}.

In order to simulate the LAr detector response, we followed Refs.~\cite{Akiri:2011dv,Agarwalla:2011hh,Coloma:2011vv}. This corresponds to an efficiency of 90\% in all signal channels (appearance and disappearance). A $0.5 \%$ neutral current events, a $1 \%$ fraction of the $\nu_\mu$ missidentified and the full intrinsic contamination of the beam were considered as backgrounds (including the oscillated events arising from the contamination with opposite polarity for each beam). The background rejection efficiencies were assumed to be constant over a neutrino energy window between 0.5 and 10~GeV. A constant Gaussian energy resolution of 150~MeV was assumed for electrons and positrons and $0.2\sqrt E$ for muons, following Ref.~\cite{Akiri:2011dv}. Migration matrices for the NC backgrounds have been kindly provided by the LBNE collaboration~\cite{lisa} and included. 

A further background component which could play a potentially important role is the decay of $\tau$ leptons. Indeed, at the oscillation peak, most of the original $\nu_\mu$ have oscillated into $\nu_\tau$. The energy of this beam is high enough so as to be above threshold for the $\nu_\tau$ charged current cross section. Thus, $\tau$ leptons will be produced and their decay products can lead to an additional background. This phenomenon, known as the $\tau$-contamination, has been studied in the context of the Neutrino Factory~\cite{Donini:2010xk,Indumathi:2009hg,Dutta:2011mc}. This background will, however, be mostly reconstructed at low energies close to the second oscillation peak. While the second peak can potentially provide very useful information, specially regarding CP violation, it is very statistically limited compared to the first peak and largely affected by neutral current backgrounds from the high energy part of the flux. Thus, in agreement with Ref.~\cite{Huber:2010dx}, we find that the physics reach of the setups studied here are not significantly affected when removing the second oscillation peak and we therefore expect no significant impact from the $\tau$-induced background. 

The following input values for the neutrino oscillation parameters have been chosen based in the most recent global analyses in
Refs.~\cite{Schwetz:2011qt,Tortola:2012te,Fogli:2012ua}: $\theta_{13}=9^\circ$, $\theta_{12}=34.2^\circ$, $\Delta m^2_{12}=7.64\times 10^{-5}$ eV$^2$,  $\theta_{23}=45^\circ$, $\Delta m^2_{31}=2.45\times 10^{-3}$ eV$^2$. All results have been obtained after marginalization over the rest of the oscillation parameters, assuming the following $1 \sigma$ Gaussian priors: $3 \%$ for $\theta_{12}$, $2.5 \%$ for $\Delta m^2_{12}$, $8 \%$ for $\theta_{23}$ and $4 \%$ for $\Delta m^2_{31}$. For $theta_{13}$ we have assumed a prior of 0.005 in $\sin^2 2 \theta_{13}$, which corresponds to the expected performance of Daya Bay once it is systematics dominated. However, we have found that this prior does not have a significant impact in any of the results shown. No prior has been assumed for $\delta$ (\textit{i.e.}, it has been left completely free during marginalization). Finally, the matter density has been computed from the PREM profile~\cite{prem}, assuming a $2\%$ uncertainty. Constant systematic uncertainties of 5 and 10\% have been assumed for the signal and background channels, respectively, for both detectors. These are fully correlated between the different bins of a particular channel and uncorrelated among the different channels. The simulation of both facilities was performed with the GLoBES software~\cite{Huber:2004ka,Huber:2007ji}.

\section{Results}
\label{sec:results}

\begin{figure}
\begin{center}
\includegraphics[width=1.\textwidth]{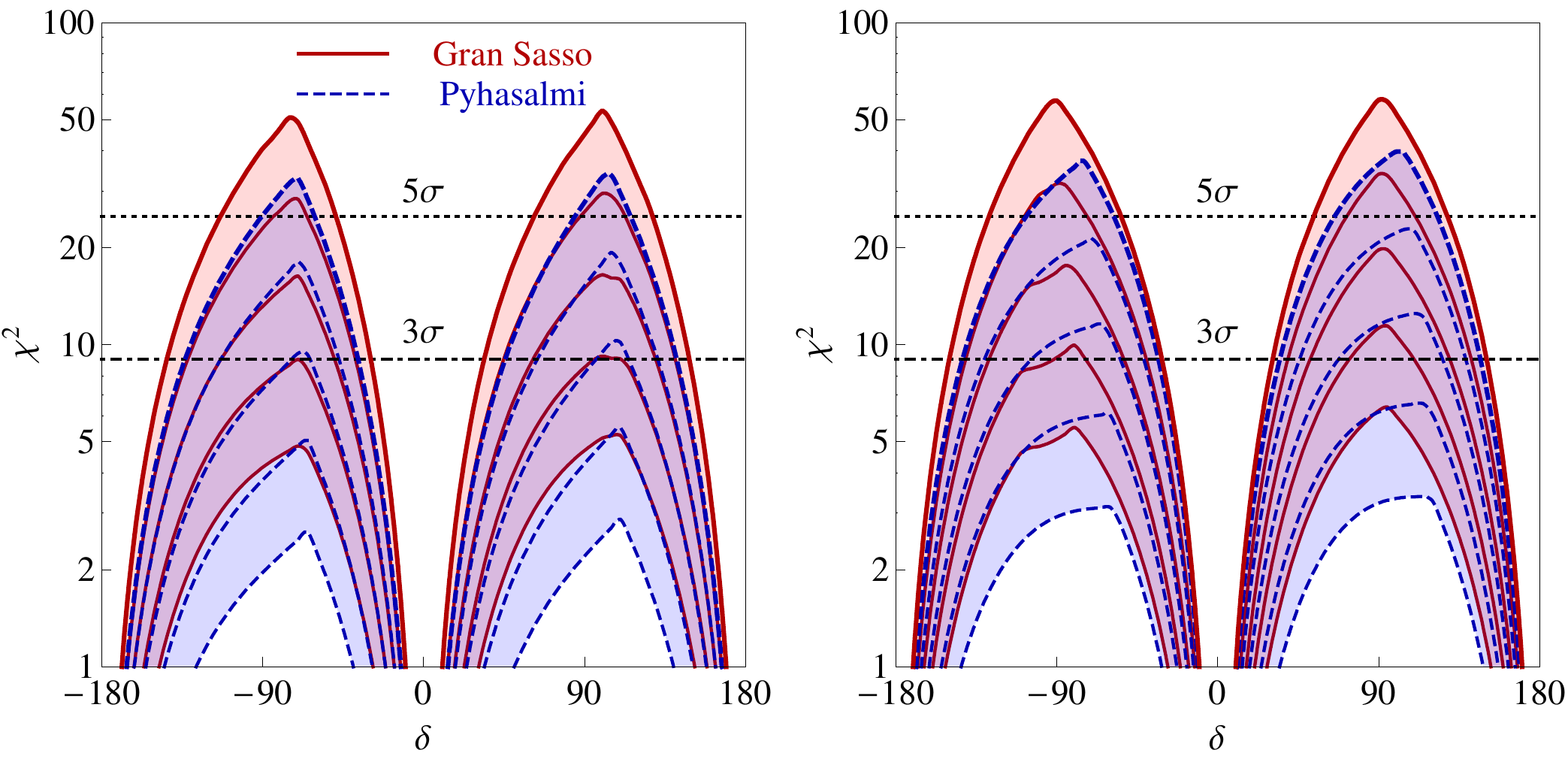} \\
\caption{Comparison of the CPV discovery potential for NH (left panel) and IH (right panel) for the two facilities under study. The top lines show the results for the maximum exposure considered for each setup, while subsequent lines show the results after reducing the statistics by factors of 2, 4, 8 and 16.}
\label{fig:CPV}
\end{center}
\end{figure}

In this section we compare the physics performance of the two scenarios under study. In Fig.~\ref{fig:CPV} we compare the CP violation (CPV) discovery potential for the two setups under study. 
Both panels show the $\chi^2$ value with which each facility would be able to disfavour CP-conservation as a function of $\delta$. The left panel shows the results for a true normal hierarchy (NH), while the right panel shows the results for inverted hierarchy (IH). Top lines correspond to the maximum exposure for each setup, while the subsequent lines in each band imply a reduction of the total exposure by factors of 2, 4, 8 and 16, to show how much a reduction of the beam power, detector mass or running time can be born without spoiling the physics performance of the facility. 
The CPV discovery potential for the two setups would thus correspond to the areas where the lines for each facility are above the corresponding value of the $\chi^2$ for a given confidence level. As an example, the 3 and 5$\sigma$ lines are shown. 

This figure shows the better performance of the shorter baseline setup combined with the WC detector for the CPV search, particularly in the case of reduced statistics. For a reduction of the statistics by more than a factor 8 neither facility has CPV discovery potential at the $3\sigma$ level. However, in the case of a factor 4 less statistics and for a NH, the shorter baseline provides 3$\sigma$ discovery potential for a $37\%$ of the possible values of $\delta$, while the longer baseline only has sensitivity for $8\%$. For the IH scenario the situation is more favourable, as we will discuss in the following, and these numbers increase to 43\% and 30\% respectively, a very remarkable improvement in the case of the longer baseline. With a factor 2 reduction in statistics the Gran Sasso setup has some sensitivity at $5\sigma$ for 12\% (21\%) of the values of $\delta$ for NH (IH), while the Pyh\"asalmi option has none. For the maximum exposures considered for each setup, the performance of the Pyh\"asalmi baseline is significantly improved to $54\%$ ($17\%$), but it is still outperformed by the shorter baseline option with $64\%$ ($37\%$) at 3$\sigma$ (5$\sigma$). This is still true for the IH case even if the Pyh\"asalmi option performs much better in this scenario, with increased sensitivity up to $62\%$ ($30\%$) of the values of $\delta$ at 3$\sigma$ (5$\sigma$) to be compared with the $66\%$ ($41\%$) that the WC option could provide.

The better performance of the WC detector for the CPV measurement is in part due to the higher statistics for the setup at Gran Sasso, which has a more massive detector. However, as can be seen in Tab.~\ref{tab:events}, the number of signal (anti)neutrino events for NH (IH) at the LAr detector is actually not very different from those at the WC. Indeed, the strong matter effects affecting the neutrino evolution will strongly enhance these channels and suppress their CP conjugate ones. This makes the event distribution between neutrinos and antineutrinos rather asymmetric at the LAr detector, negatively affecting its sensitivity to CPV. This unbalance is much less pronounced in the IH case. In this case, while matter effects tend to suppress the neutrino oscillation probability, the higher neutrino fluxes and cross sections with respect to the antineutrino ones ensure a more symmetric distribution of the events, leading to a remarkable enhancement in sensitivity to CPV for the LAr alternative, as discussed in Fig.~\ref{fig:CPV}.

\begin{table}[htb]
\begin{center}{
\renewcommand{\arraystretch}{1.5}
 \begin{tabular}{|c|c|c c|c c|}
\cline{3-6}
\multicolumn{1}{c}{ } &\multicolumn{1}{c}{ } & \multicolumn{2}{|l|}{\;500 kt WC; $L=730$ km\; } & \multicolumn{2}{|c|}{\;100 kt LAr; $L=2300$ km\; } \\ \hline
\; Hierarchy \;     & \; $\delta$ \; & \qquad $N_\nu$\; & \;$N_{\bar{\nu}}$\; & \qquad $N_\nu$\; & \;$N_{\bar{\nu}}$\;  \\ \hline \hline
\multirow{3}{*}{NH} & $ \pi/2$ \; 	& \qquad  1526 & 931 & \qquad 1374 & 230 \cr \cline{2-6}
		    & $ 0$ 		& \qquad  2012 & 782 & \qquad 1650 & 195 \cr \cline{2-6}
		    & $ -\pi/2$\; 	& \qquad  2464 & 546 & \qquad 1989 & 113 \\ \hline
\multirow{3}{*}{IH} & $ \pi/2$\; 	& \qquad  965 & 1329 & \qquad 250 & 884 \cr \cline{2-6}
		    & $ 0$ 		& \qquad  1296 & 1074 & \qquad 338 & 769 \cr \cline{2-6}
		    & $ -\pi/2$\; 	& \qquad  1717 & 852 & \qquad 525 & 621 \\ \hline \hline
\multicolumn{2}{|c|}{Background} 				&  \qquad 1094 & 743 & \qquad 264 & 120 \\ \hline 
 \end{tabular}}
\caption{Number of $\nu_e$ and $\bar{\nu}_e$ events at the 500~kt fiducial WC detector located at 730~km and at the 100~kt LAr detector placed at 2300~km. Signal events are show for normal (NH) and inverted (IH) mass hierarchy and for three values of $\delta$ to show the dependence on these two parameters. Background events are shown for $\delta=0$ and NH. The values for the rest of the oscillation parameters are the ones listed in Sec.~\ref{sec:setups}.
\label{tab:events} }
\end{center}
\end{table}

\begin{figure}
\begin{center}
\includegraphics[width=1.\textwidth]{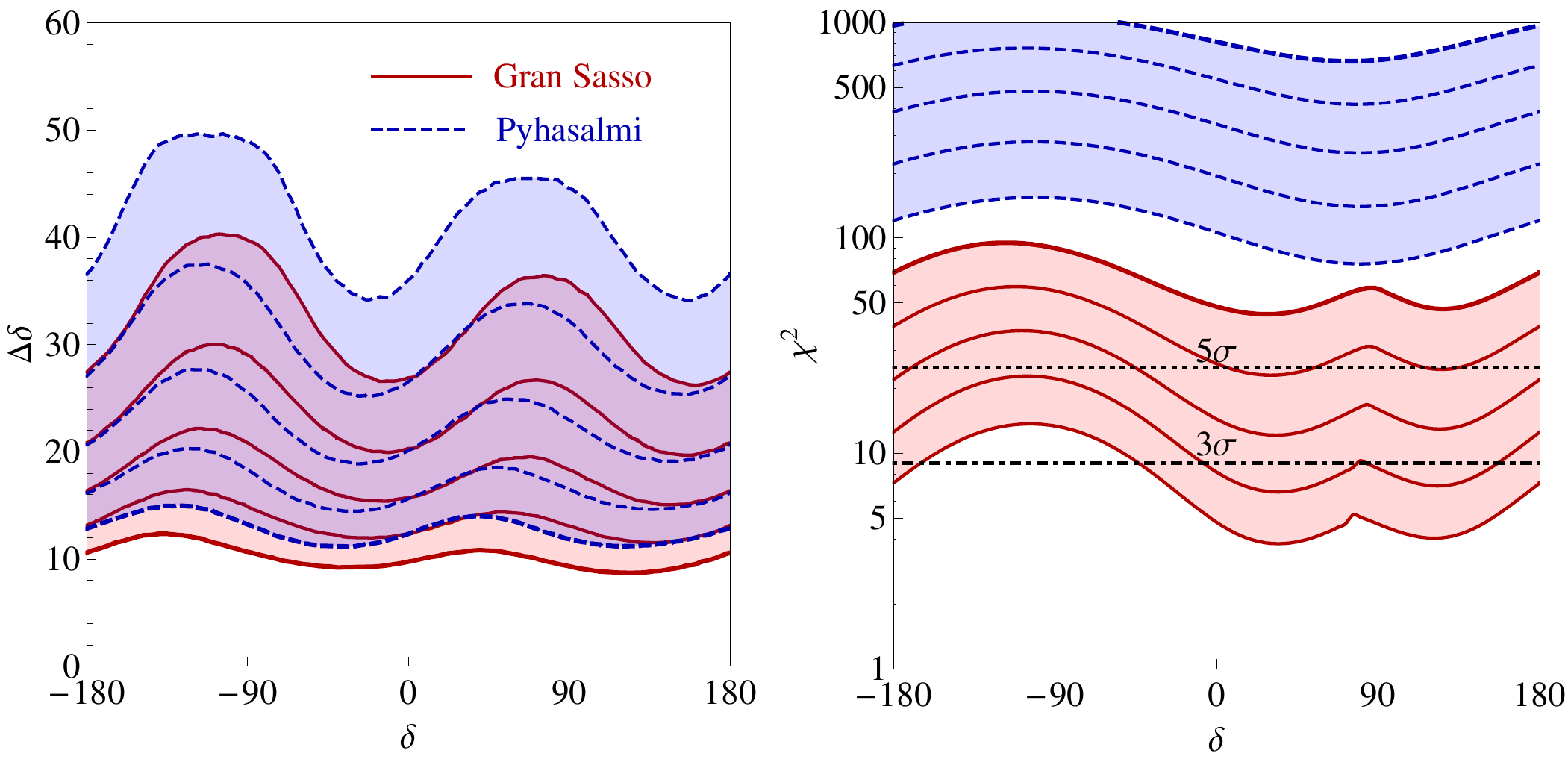} \\
\caption{Comparison of the achievable precision at $1\sigma$ in $\delta$ (left panel) and the mass hierarchy discovery potential (right panel) for the two facilities under study. In the left (right) panels, the bottom (top) lines show the results for the maximum exposure considered for each setup, while subsequent lines show the results after reducing the statistics by factors of 2, 4, 8 and 16. A normal hierarchy has been assumed. }
\label{fig:prec}
\end{center}
\end{figure}

Fig.~\ref{fig:prec} shows the results for the achievable precision on $\delta$ and the mass hierarchy discovery potential. A normal hierarchy has been assumed in this case for both panels.  
In the left (right) panels, the bottom (top) lines show the results for the maximum exposure considered, while subsequent lines show the results after reducing the statistics by factors of 2, 4, 8 and 16. The left panel shows $\Delta \delta$, defined as $1/2$ of the 1$\sigma$ allowed region in the measurement of the CP violating phase $\delta$. We plot this as a function of the true value of $\delta$ since this dependence is quite strong for this observable (see Ref.~\cite{Coloma:2012wq} for a detailed study). We find that the 730~km option consistently performs better in this measurement than the 2300~km option. Furthermore, when reducing the statistics the deterioration of the Pyh\"asalmi setup in this measurement is faster than for the shorter baseline option, as shown by the more widely spaced lines. 

The mass hierarchy discovery potential is depicted in the right panel in Fig.~\ref{fig:prec} for both setups under study. In this case, the $\chi^2$ value with which each facility can disfavour the wrong mass hierarchy is shown as a function of the true value of $\delta$. We only show the results assuming a true normal hierarchy in this case: the results for inverted hierarchy are very similar to these under the inversion $\delta \rightarrow -\delta$. In this measurement, the much stronger matter effects at the Pyh\"asalmi baseline would allow to perform this measurement with much smaller exposure, clearly outperforming the shorter baseline by far. However, a $5\sigma$ determination of the mass hierarchy at the shorter baseline is still possible for any value of $\delta$. Even reducing the statistics by a factor two the $5 \sigma$ level can be reached in almost all the parameter space. Thus, the short baseline seems also adequate to perform this measurement if no higher significance is required.

\begin{figure}
\begin{center}
\includegraphics[width=1.\textwidth]{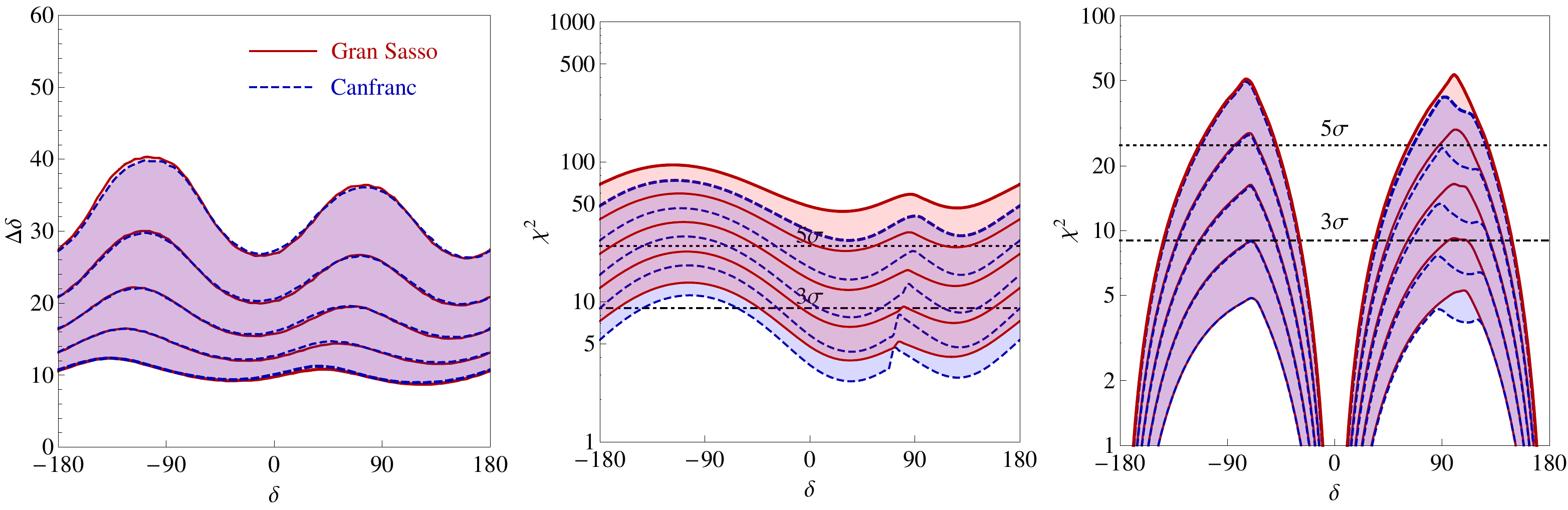} \\
\caption{Comparison of $\Delta \delta$ (left panel), the mass hierarchy (middle panel) and the CPV discovery potential (right panel) for the WC detector placed at Gran Sasso or Canfranc. For all observables, the best results correspond to the maximum exposure considered while the different lines show the results after reducing the statistics by factors of 2, 4, 8 and 16.}
\label{fig:sensitivitiescf}
\end{center}
\end{figure}

It is well-known that the size of the Gran Sasso underground laboratory is physically limited and therefore it may have difficulties in hosting a very massive detector. As it can be seen from Figs.~\ref{fig:CPV} and~\ref{fig:prec}, the physics reach for this setup after a reduction of the detector mass by a factor of 2 is still quite good, though. However, if a WC detector of the desired volume cannot be accommodated at Gran Sasso, the closest alternative options are Canfranc (650~km) and Umbria (665~km). 
In Fig.~\ref{fig:sensitivitiescf} we compare the same performance indicators as in Figs.~\ref{fig:CPV} and~\ref{fig:prec} but for the Gran Sasso and Canfranc baselines. As expected, the $1\sigma$ precision on the measurement of $\delta$ is unaffected since at $1\sigma$ the mass hierarchy is always solved. The mass hierarchy discovery potential is, however, affected due to the smaller matter effects at this baseline, and an overall reduction of the significance for which the wrong hierarchy can be ruled out takes place for all values of $\delta$. 
This in turn limits the CPV discovery potential since the sign degeneracy can mimic CP conservation for some CP-violating values of $\delta$. This loss in sensitivity can be seen in Fig.~\ref{fig:sensitivitiescf} for a small area around $\delta = \pi/2$ for NH. A similar effect takes place around $\delta = -\pi/2$ for IH. Apart from this small region, the rest of the CPV discovery potential is unaffected by the change in baseline. It should also be noted that in the full statistics scenario the mass hierarchy can still be determined at $5\sigma$ for any value of $\delta$. Thus, while the Gran Sasso baseline provides a better physics reach, the Canfranc alternative always performs very similarly, providing a reasonable compromise.

\section{Summary and discussion}
\label{sec:summary}

In this work we have studied the physics performance of a long baseline neutrino oscillation experiment based on a neutrino beam from the CERN accelerator complex. In general, for a fixed beam power, very energetic neutrino beams aiming to correspondingly long baselines (so that $L/E$ is close to the first oscillation peak) tend to give the best performance. Indeed, while longer baselines imply a flux reduction with $L^{-2}$, the linear increase of the neutrino cross section with the energy and the higher focusing of the beam at high energies lead to an overall increase in statistics which is more or less linear in energy. However, the detector response can greatly vary at different energies, depending on the chosen detector technology. The detection technology also determines the maximum mass that can be reached in each case. 

In this work we have compared the physics reach of two representative setups for a SuperBeam in Europe. These correspond to very different baselines, and therefore make use of the different detector technologies which better match their needs. The very good CPV discovery potential of the SPL ($L=130$ km) is very well-known and has been widely studied in the literature. However, it has no sensitivity to the mass hierarchy from long baseline oscillations alone, and relies on the availability of a 4~MW beam which is not expected to be at hand in the near future. Therefore, longer baselines have been considered for all the setups presented in this work, and reduced beam powers of 0.8 MW. We considered 5 years data taking with each beam polarity. 

On one hand, we have considered a baseline around $600-700$ km. Several possible underground laboratories match this baseline from CERN. The Gran Sasso laboratory, placed at $L=730$~km, has the advantage of an existing beamline from CERN. An interesting alternative would be offered by the existing underground laboratory at Canfranc ($L=650$ km), one of the seven LAGUNA candidate sites. In order to match the first oscillation peak at these baselines, the neutrino flux should be peaked around 1-2~GeV. It is well-known that WC detectors perform optimally for neutrino energies precisely in this range, where the QE cross section peaks. They can also be built on very large scales. Therefore, we have considered a 500 kt WC for these baselines. On the other hand, we have also considered a setup with a very long baseline ($L=2300$ km). This matches the distance from CERN to Pyh\"asalmi, which is also one of the considered sites within LAGUNA. In this case, the neutrino flux should be peaked around 4-5 GeV. Consequently, most of the events would lie in the deep inelastic scattering regime where the WC detector is no longer optimal due to its poor efficiency for multi-ring events. Instead, LAr constitutes an ideal detector technology for this setup, with very high efficiencies and extremely good energy resolution for deep inelastic events. We have considered a maximum fiducial mass of 100 kt for this detector instead.

We have compared the physics reach of these three setups (WC at 730 and 650~km, and LAr at 2300~km) under three different performance indicators: the precision in their measurement of the CP violating phase $\delta$, their CPV discovery potential and their mass hierarchy discovery potential. For each of these indicators we have also studied scenarios with reduced statistics to explore if reductions of the beam power, detector mass and/or running times are possible given the large value of $\theta_{13}$ recently discovered. 

We find that, for the mass hierarchy discovery potential, the much stronger matter effects at the 2300~km LAr option greatly outperform the shorter baselines. Indeed, a $\sim 10 \sigma$ exclusion of the wrong hierarchy can be accomplished even with a reduction of the statistics by a factor of 16. However, the shorter baselines can also provide an adequate determination of the mass hierarchy. Indeed, the Canfranc option can rule out the wrong mass hierarchy at $5 \sigma$ with the maximum exposure considered. The situation is slightly better for the setup at Gran Sasso, which due to its slightly larger matter effects can do it even after a reduction of statistics by almost a factor 2. 

Regarding the measurement of the CP violating phase, the short baselines are clearly preferable. Indeed, the strong matter effect enhancement of the oscillation probability at high energies also leads to a very asymmetric distribution of the events between the two beam polarities and to a reduction of their dependence on $\delta$, deteriorating its measurement. We find that the setups with shorter baselines can provide a measurement of $\delta$ with an error at $1\sigma$ which ranges from $10^\circ$ to $12^\circ$ depending on the value of $\delta$, while for the setup at Pyh\"asalmi this ranges between $12^\circ$ to $15^\circ$ for the maximum exposure scenario. As the statistics is reduced, the measurement is deteriorated faster for the detector placed at $L=2300$ km. For example, if the exposure is reduced by a factor 8 the $1 \sigma$ error on delta would be between $20^\circ-30^\circ$ for WC, while for the LAr option it would be around $26^\circ-38^\circ$. Similarly, the CPV discovery potential is better at the shorter baselines. At Gran Sasso, CPV could be found for a $64\%$ ($37\%$) of the possible values of $\delta$ at 3$\sigma$ (5$\sigma$). These numbers are reduced to $54\%$ ($17\%$) for LAr at the longer baseline. For the more favourable inverted hierarchy scenario the coverage of both facilities improves, specially for the LAr option, reaching $66\%$ ($41\%$) for the WC and $62\%$ ($30\%$) for LAr.   

\section{Conclusions}
\label{sec:concl}    

We conclude that a long baseline neutrino oscillation experiment with a CERN-produced beam aiming to a large detector in an underground laboratory in Europe can provide excellent sensitivities to the two remaining unknowns among the neutrino oscillation parameters: the mass hierarchy and the existence of leptonic CPV. A liquid argon (LAr) detector placed at a very long baseline would grant an exceptional discovery potential to the mass hierarchy through its strong matter effects. On the other hand, the very same matter effects limit its sensitivity to $\delta$. The opposite is true for a water \v{C}erenkov (WC) detector placed at a shorter baseline. The smaller matter effects translate in an enhanced sensitivity to $\delta$ but a much poorer mass hierarchy discovery potential.

The ability to observe CP violation (CPV) in a large fraction of the parameter space is related to the precision which can be achieved for a measurement of $\delta$. It is therefore desirable to maximize it, given that it is not possible to be sensitive to CPV in the whole parameter space. 
Regarding the mass hierarchy discovery potential, on the other hand, once the desired confidence level has been reached a more accurate measurement is not particularly helpful, since it is a discrete parameter.
For these reasons we conclude that, under the assumptions made for the simulation of each detector (detailed in Section~\ref{sec:setups}), the shorter baseline options combined with a WC detector are generally preferable, since they reach a better precision in the measurement of $\delta$ and hence provide a larger coverage for CPV while they still achieve $5 \sigma$ sensitivity to the mass hierarchy for any value of $\delta$.

\begin{acknowledgments}

This work was partially funded by the European Community under the European Commission Framework Programme 7 Research Infrastructure Design Studies EUROnu (212372, FP7-INFRA-2007-1), LAGUNA (212343, FP7-INFRA-2007-1), LAGUNA-LBNO (Project Number 284518, FP7-INFRA-2007-1) and the ITN INVISIBLES (Marie Curie Actions, PITN-GA-2011-289442). PC has been supported by the U.S. Department of Energy under award number DE-SC0003915. The views expressed are not necessarily those of the funding bodies.

\end{acknowledgments}


\end{document}